\documentstyle[aps]{revtex}
\begin{document}
\voffset 0.8truecm
\title{
Quantum cloning machines for equatorial qubits
}
\author{
Heng Fan$^a$, Keiji Matsumoto$^a$, Xiang-Bin Wang$^a$,
and Miki Wadati$^b$}
\address{
$^a$Imai quantum computing and information project,
ERATO, \\
Japan Science and Technology Corporation,\\
Daini Hongo White Bldg.201, Hongo 5-28-3, Bunkyo-ku, Tokyo 133-0033, Japan.\\
$^b$Department of Physics, Graduate School of Science,\\
University of Tokyo, Hongo 7-3-1, Bunkyo-ku, Tokyo 113-0033, Japan.
}
\maketitle
                                                        
\begin{abstract}
Quantum cloning machines for equatorial qubits are studied.
For the case of 1 to 2 phase-covariant quantum cloning machine,
we present the networks consisting of quantum gates
to realize the quantum cloning
transformations. The copied equatorial qubits are shown
to be separable by using Peres-Horodecki criterion.
The optimal 1 to $M$ phase-covariant quantum cloning
transformations are given.
\end{abstract}              
       

\newpage
\section{Introduction}
Quantum computing and quantum information have been attracting a great
deal of interests. They differ in many aspects from the classical
theories. One of the most fundamental differences between classical and
quantum information is the no-cloning theorem\cite{WZ}. It tells us that
an arbitrary quantum information can not be copied exactly.
The no-cloning theorem for pure states is also extended to the case
that a general mixed state can not be broadcast\cite{BCFJ}. 
However, no-cloning theorem does not forbid imperfect cloning.
And several kinds of quantum cloning machines (QCM)
are proposed, the optimal fidelity and transformations
of QCMs are found
\cite{BH,BDEF,F,GM,BEM,W,KW}.

In the proof of the no-cloning theorem, Wootters and Zurek introduced
a QCM which has the property that the
quality of the copy it makes depends on the input states\cite{WZ}.
To diminish or cancel this disadvantage, Bu\v{z}ek and Hillery proposed a
universal quantum cloning machine (UQCM) for an arbitrary pure state
where the copying process is input-state independent.
They use Hilbert-Schmidt norm to quantify distances between
input density operator and the output density operators.
Bru\ss ~et al \cite{BDEF} discussed the performance of a UQCM by
analyzing the role of the symmetry and isotropy conditions
imposed on the system and found the optimal UQCM and the
optimal state-dependent quantum cloning.
Optimal fidelity and optimal quantum cloning
transformations of general $N$ to $M$ ($M>N$) case are presented
in Ref.\cite{GM,BEM,W,KW}.
The relation between quantum cloning and superluminal
signalling is proposed and discussed in Ref.\cite{G,BDMS}.
It was also shown that the UQCM can be realized by 
a network consisting of quantum gates\cite{BBHB}.

In case of UQCM, the input states are arbitrary pure states.
In this paper, we study the QCM for a restricted set of pure input
states. The Bloch vector is restricted to the intersection
of $x-z$ ($x-y$ and $y-z$) plane with the Bloch sphere. This
kind of qubits are the so-called equatorial qubits\cite{BCDM}
and the corresponding QCM is called phase-covariant quantum
cloning.
The 1 to 2 phase-covariant quantum cloning was first studied 
by Bru\ss ~et al \cite{BCDM} who studied the
optimal quantum cloning for $x-z$ equatorial qubits by taking
BB84 states as input.
The fidelity of quantum cloning for the equatorial qubits
is higher than the original Bu\v{z}ek and Hillery UQCM\cite{BH}.
This is expected as the more information about the input
is given, the better one can clone each of its states.

In this paper, using the approach presented in Ref.\cite{BBHB}, we show
that the 1 to 2 optimal phase-covariant quantum cloning machines
can be realized by networks consisting of quantum rotation
gates and controlled NOT gates. The copied equatorial
qubits are shown to be separable by using Peres-Horodecki
criterion\cite{P,H}. We then present the 1 to $M$ phase-covariant
quantum cloning transformations and prove that the
fidelity is optimal. The general $N$ to $M$ ($M>N$) optimal
phase-covariant quantum cloning machines are finally
proposed.

The paper is organized as follows: In section 2,
we introduce the cloning transformations for equators in
$x-z$ and $x-y$ planes.
In section 3, phase-covariant quantum
cloning can be realized by networks consisting of
quantum gates. In section 4, the copied qubits
are shown to be separable and quantum triplicators
are studied. In section 5, optimal 1 to $M$ phase-covariant
quantum cloning machines are presented and proved.
In section 6, $N$ to $M$ ($M>N$) phase-covariant QCM
is proposed.
Section 7 includes a brief summary.

\section{1 to 2 phase-covariant quantum cloning}
Instead of arbitrary input states,
we consider the input state which we intend to clone
to be a restricted set of states.
It is a pure superposition state 
\begin{eqnarray}
|\Psi \rangle =\alpha |0\rangle +\beta |1\rangle 
\label{input}
\end{eqnarray}
with $\alpha ^2+\beta ^2=1$. Here, we use an assumption that
$\alpha $ and $\beta $ are real in contrast to
complex when we consider the case of UQCM.
That means the $y$ component
of the Bloch vector of the input qubit is zero.
Because that there are just one unknown parameter in the input state
under consideration, we expect that we can
achieve a better quality in quantum cloning if we can find
an appropriate phase-covariant QCM.

The case of 1 to 2 phase-covariant quantum cloning transformation
has already been found by Bru\ss ~et al in \cite{BCDM}.
They proposed the following cloning transformation
for the input (\ref{input}),
\begin{eqnarray}
|0\rangle _{a_1}|Q\rangle _{a_2a_3}&\rightarrow &
[({1\over 2}+\sqrt{1\over 8})|00\rangle _{a_1a_2}+
({1\over 2}-\sqrt{1\over 8})
|11\rangle _{a_1a_2}]|\uparrow \rangle _{a_3}
+\frac 12|+\rangle _{a_1a_2}|\downarrow \rangle _{a_3},
\noindent \\
|1\rangle _{a_1}|Q\rangle _{a_2a_3}&\rightarrow &
[({1\over 2}+\sqrt{1\over 8})|11\rangle _{a_1a_2}+
({1\over 2}-\sqrt{1\over 8})
|00\rangle _{a_1a_2}]|\downarrow \rangle _{a_3}
+\frac 12|+\rangle _{a_1a_2}|\uparrow \rangle _{a_3},
\label{optimal}
\end{eqnarray}
where the following notations are introduced
\begin{eqnarray}
|+\rangle =\frac 1{\sqrt{2}}(|10\rangle +|01\rangle ), ~~
|-\rangle =\frac 1{\sqrt{2}}(|10\rangle -|01\rangle ).
\label{base}
\end{eqnarray}
The fidelity of the phase-covariant cloning transformation is
$F={1\over 2}+\sqrt{1\over 8}$ which is larger than $F={5\over 6}$,
the fidelity of 1 to 2 UQCM\cite{BH}. And also this fidelity was
proved to be
optimal for phase-covariant cloning machine\cite{BCDM}.
Actually, because we assume $\alpha $ and $\beta $ are real,
only a single unknown parameter is copied instead of two
unknown parameters for the case of a general pure state.
Thus a higher fidelity of quantum cloning can be achieved.
The case of spin flip has a similar phenomenon\cite{BHW,BBHB,B}. 
Here, the fidelity is defined in the standard form as
$F=\langle \Psi |\rho |\Psi \rangle $, $\rho $ is the output
reduced density operator at a single qubit.

For convenience, we present the following cloning transformation
for pure input state (\ref{input}),
\begin{eqnarray}
|0\rangle _{a_1}|Q\rangle _{a_2a_3}\rightarrow
\left( |00\rangle _{a_1a_2}
+\lambda |11\rangle _{a_1a_2}\right)q|\uparrow \rangle _{a_3}
+\left( |10\rangle _{a_1a_2}
+|01\rangle _{a_1a_2}\right) y|\downarrow \rangle _{a_3}, \nonumber \\
|1\rangle _{a_1}|Q\rangle _{a_2a_3}\rightarrow
\left( |11\rangle _{a_1a_2}
+\lambda |00\rangle _{a_1a_2}\right) q|\downarrow \rangle _{a_3}
+\left( |10\rangle _{a_1a_2}
+|01\rangle _{a_1a_2}\right) y|\downarrow \rangle _{a_3},
\label{tran2}
\end{eqnarray}
where we assume $\lambda $ is real and $\lambda \not= \pm 1$,
we also use notations
\begin{eqnarray}
q\equiv \sqrt{\frac {2}{3-2\lambda +3\lambda ^2}},
~~~y\equiv \frac {1-\lambda }{\sqrt{6-4\lambda +6\lambda ^2}}.
\label{qy}
\end{eqnarray}
The qubit in $a_1$ is the input state, the output copies appear
in $a_1, a_2$ qubits, and $a_3$ is the ancilla state.
In case $\lambda =0$, the cloning transformation reduces to
the UQCM proposed in \cite{BH}.
When $\lambda =3-2\sqrt{2}$, we obtain the optimal phase-covariant
quantum cloning transformation presented in \cite{BCDM} for
$x-z$ equator.
Actually, we can use both Bures fidelity and Hilbert-Schmidt norm
to quantify the quality of the copies\cite{KOWY}. Both of them show that
transformation (\ref{optimal}) is the optimal cloning machine
for input state (\ref{input}).

Sometimes, we study $x-y$ equator instead of $x-z$ equator so that
some results can be obtained easier, and the two cases are
connected by a transformation. We consider the input state
as
\begin{eqnarray}
|\Psi \rangle &=&\frac {1}{\sqrt{2}}[|0\rangle +e^{i\phi }
|1\rangle ],
\label{xy}
\end{eqnarray}
where $\phi \in [0,2\pi )$.
One can check that the $y$ component of the
Bloch vector of this state is zero.
The cloning transformation takes the form,
\begin{eqnarray}
|0\rangle _{a_1}|00\rangle _{a_2a_3}\rightarrow
\frac {2(1-\lambda )}{\sqrt{6-4\lambda +6\lambda ^2}}|00\rangle _{a_1a_2}
|0\rangle _{a_3} +
\frac {1+\lambda }{\sqrt{6-4\lambda +6\lambda ^2}}
\left( |01\rangle _{a_1a_2}+|10\rangle _{a_1a_2}\right)
|1\rangle _{a_3},
\nonumber \\
|1\rangle _{a_1}|00\rangle _{a_2a_3}\rightarrow
\frac {2(1-\lambda )}{\sqrt{6-4\lambda +6\lambda ^2}}|11\rangle _{a_1a_2}
|1\rangle _{a_3} +
\frac {1+\lambda }{\sqrt{6-4\lambda +6\lambda ^2}}
\left( |01\rangle _{a_1a_2}+|10\rangle _{a_1a_2}\right)
|0\rangle _{a_3}.
\label{xytran}
\end{eqnarray}
As the case of $x-y$ equator, $\lambda =0$ corresponds to UQCM,
and the case $\lambda =3-2\sqrt {2}$ is the optimal phase-covariant
quantum cloning for input (\ref{xy}) which takes
the following form
\begin{eqnarray}
|0\rangle _{a_1}|00\rangle _{a_2a_3}\rightarrow
\frac {1}{\sqrt{2}}|00\rangle _{a_1a_2}|0\rangle _{a_3}
+{\frac 12}\left( |01\rangle _{a_1a_2}+
|10\rangle _{a_1a_2}\right) |1\rangle _{a_3},
\nonumber \\
|1\rangle _{a_1}|00\rangle _{a_2a_3}\rightarrow
\frac {1}{\sqrt{2}}|11\rangle _{a_1a_2}|1\rangle _{a_3}
+{\frac 12}\left( |01\rangle _{a_1a_2}+
|10\rangle _{a_1a_2}\right) |0\rangle _{a_3}.
\label{xyopt}
\end{eqnarray}

\section{Quantum cloning networks for equatorial qubits}
In this section, following the method proposed by
Bu\v{z}ek $et ~al$\cite{BBHB}, we show that the quantum cloning
transformations for equatorial qubits can be realized by
networks consisting of quantum logic gates.
Let us first introduce the method proposed by Bu\v{z}ek $et~al$\cite{BBHB},
and then analyze the case of phase-covariant
cloning.
The network is
constructed by one- and two-qubit gates. The
one-qubit gate is a single qubit rotation operator
$\hat {R}_j(\vartheta )$,
defined as
\begin{eqnarray}
\hat {R}_j(\vartheta )|0\rangle _j&=&\cos \vartheta |0\rangle _j
+\sin \vartheta |1\rangle _j,\nonumber \\
\hat {R}_j(\vartheta )|1\rangle _j&=&
-\sin \vartheta |0\rangle _j+\cos \vartheta |1\rangle _j.
\end{eqnarray}
The two-qubit gate is the controlled NOT gate represented
by the unitary matrix
\begin{eqnarray}
\hat{P}=\left( \begin{array}{cccc}
1&0&0&0\\
0&1&0&0\\
0&0&0&1\\
0&0&1&0\end{array}\right) . 
\end{eqnarray}
Explicitly, the controlled NOT gate
$\hat {P}_{kl}$ acts on the basis vectors of the two qubits
as follows:
\begin{eqnarray}
\hat{P}_{kl}|0\rangle _k|0\rangle _l=|0\rangle _k|0\rangle _l,
~~\hat{P}_{kl}|0\rangle _k|1\rangle _l=|0\rangle _k|1\rangle _l,
\nonumber \\
\hat{P}_{kl}|1\rangle _k|0\rangle _l=|1\rangle _k|1\rangle _l,
~~\hat{P}_{kl}|1\rangle _k|1\rangle _l=|1\rangle _k|0\rangle _l.
\end{eqnarray}
Due to Bu\v{z}ek $et~al$, the action of the copier is expressed as
a sequence of two unitary transformations,
\begin{eqnarray}
|\Psi _{a_1}^{(in)}|0\rangle _{a_2}|0\rangle _{a_3}
\rightarrow |\Psi \rangle _{a_1}^{(in)}|\Psi \rangle _{a_1a_2}^{(prep)}
\rightarrow |\Psi \rangle _{a_1a_2a_3}^{(out)}.
\end{eqnarray}
This network can be described by a figure in Ref.\cite{BBHB}.
The preparation state is constructed as
\begin{eqnarray}
|\Psi \rangle _{a_2a_3}^{(prep)}=\hat {R}_2(\vartheta _3)
\hat {P}_{32}\hat {R}_3(\vartheta _2)\hat {P}_{23}
\hat {R}_2(\vartheta _1)|0\rangle _{a_2}|0\rangle _{a_3}.
\end{eqnarray}
The quantum copying is performed by
\begin{eqnarray}
|\Psi \rangle _{a_1a_2a_2}^{(out)}=\hat {P}_{a_3a_1}\hat {P}_{a_2a_1}
\hat {P}_{a_1a_3}\hat {P}_{a_1a_2}|\Psi \rangle _{a_1}^{(in)}
|\Psi \rangle _{a_2a_3}^{(prep)}.
\label{copy}
\end{eqnarray}
Note that the output copies appear in the $a_2, a_3$ qubits instead
of $a_1, a_2$ qubits. For UQCM, we should choose\cite{BBHB}
\begin{eqnarray}
\vartheta _1=\vartheta _3=\frac {\pi }{8},
~~\vartheta _2=-\arcsin \left( \frac {1}{2}-\frac {\sqrt{2}}{3}\right) ^{1/2}.
\end{eqnarray}

We now consider the cloning transformations for equatorial qubits.
The network proposed by Bu\v{z}ek $et~al$ is rather general.
We only need to take a different angles $\vartheta _j,j=1,2,3$
to realize the phase-covariant cloning.
In the case of cloning transformation for $x-y$ equator (\ref{xytran}),
the preparation state takes the form
\begin{eqnarray}
|\Psi \rangle _{a_2a_3}^{(perp)}=
\frac {2(1-\lambda )}{\sqrt{6-4\lambda +6\lambda ^2}}|00\rangle _{a_2a_3}
+
\frac {1+\lambda }{\sqrt{6-4\lambda +6\lambda ^2}}
\left( |01\rangle _{a_1a_2}+|10\rangle _{a_2a_3}\right) .
\end{eqnarray}
The preparation state corresponding to cloning transformation
(\ref{tran2}) for $x-z$ equator can be written as
\begin{eqnarray}
|\Psi \rangle _{a_2a_3}^{(perp)}=
q|00\rangle _{a_2a_3}+q\lambda |11\rangle _{a_2a_3}+y|10\rangle _{a_2a_3}+y|01\rangle _{a_2a_3}.
\end{eqnarray}
We can check that for some angles $\vartheta _j,j=1,2,3$,
the above preparation states can be realized,
Actually we have several choices. When $\lambda =0$,
we obtain the result for UQCM.
Here we present the
result for the optimal case, i.e., $\lambda =3-2\sqrt{2}$.

For $x-y$ equator, let
\begin{eqnarray}
\vartheta _1=\vartheta _3=\arcsin \left( \frac {1}{2}
-\frac {1}{2\sqrt{3}}\right)^{1\over 2},
~~\vartheta _2=-\arcsin \left(
\frac {1}{2}-\frac {\sqrt{3}}{4}\right) ^{1\over 2}.
\end{eqnarray}
Then, the preparation state has the form
\begin{eqnarray}
|\Psi \rangle _{a_2a_3}^{(perp)}
=\frac {1}{\sqrt{2}}|00\rangle _{a_2a_3}
+\frac {1}{2}(|01\rangle _{a_2a_3}+|10\rangle _{a_2a_3}).
\end{eqnarray}
For $x-z$ equator, let
\begin{eqnarray}
\vartheta _1=\vartheta _3=\arcsin \left( \frac {1}{2}
-\sqrt{\frac {1}{8}}\right)^{1\over 2},
~~\vartheta _2=0.
\end{eqnarray}
The preparation state is
\begin{eqnarray}
|\Psi \rangle _{a_2a_3}^{(perp)}
=\left( \frac {1}{2}+\sqrt{\frac {1}{8}}\right) |00\rangle _{a_2a_3}
+\frac {1}{2\sqrt{2}}(|01\rangle _{a_2a_3}+|10\rangle _{a_2a_3})
+\left( \frac {1}{2}-\sqrt{\frac {1}{8}}\right) |11\rangle _{a_2a_3}.
\end{eqnarray}
After the preparation stage,
perform the copying procedure (\ref{copy}), we obtain the
output state. And the output copies appear in the
$a_2$ and $a_3$ qubits. The optimal quantum cloning
transformations for equatorial qubits can achieve the highest fidelity
$\frac {1}{2}+\sqrt{\frac {1}{8}}$.
The reduced density operator of both copies at the output in
$a_2$ and $a_3$ qubits
can be expressed as
\begin{eqnarray}
\rho ^{(out)}=\left( \frac {1}{2}+\sqrt{\frac {1}{8}}\right)
|\Psi \rangle \langle \Psi |+\left( \frac {1}{2}-\sqrt{\frac {1}{8}}\right)
|\Psi _{\perp}\rangle \langle \Psi _{\perp}|.
\end{eqnarray}

\section{Separability of copied qubits and quantum
triplicators}
\subsection{Separability}
For the UQCM, the density matrix for the two copies
$\rho _{a_2a_3}^{(out)}$ is shown to be inseparable
by use of Peres-Horodecki criterion\cite{P,H}. That means it cannot
be written as the convex sum,
\begin{eqnarray}
\rho _{a_2a_3}^{(out)}=\sum _mw^{(m)}\rho _{a_2}^{(m)}\otimes
\rho _{a_3}^{(m)},
\end{eqnarray}
where the positive weights $w^{(m)}$ satisfy $\sum _mw^{(m)}=1$.
And there are correlations between the copies, i.e., the two qubits
at the output of the quantum copier are nonclassically
entangled\cite{BBHB}.
We shall show in this section that,
different from the UQCM, the copied qubits are
separable for the case of optimal phase-covariant quantum cloning
by Peres-Horodecki criterion.

Peres-Horodecki's positive partial transposition criterion
states that
the positivity of the partial transposition of a state
is both necessary and sufficient condition for its
separability\cite{P,H}.
For $x-z$ equator where the input state is $\alpha |0\rangle +\beta |1\rangle $,
with $\alpha =\cos \theta ,\beta =\sin \theta $,
the partially transposed output
density operator at $a_2, a_3$ qubits is expressed by
a matrix,
\begin{eqnarray}
[\rho _{a_2a_3}^{(out)}]^{T_2}=
\frac {1}{3-2\lambda +3\lambda ^2}\left(
\begin{array}{cccc}
2(\alpha ^2+\lambda ^2\beta ^2)
&\alpha \beta (1-\lambda ^2)
&\alpha \beta (1-\lambda ^2)
&\frac {1}{2}(1-\lambda )^2\\
\alpha \beta (1-\lambda ^2)
&\frac {1}{2}(1-\lambda )^2 & 2\lambda
&\alpha \beta (1-\lambda ^2)\\
\alpha \beta (1-\lambda ^2)
&2\lambda &\frac {1}{2}(1-\lambda )^2
&\alpha \beta (1-\lambda ^2)\\
\frac {1}{2}(1-\lambda )^2 
&\alpha \beta (1-\lambda ^2)
&\alpha \beta (1-\lambda ^2)
&2(\beta ^2+\alpha ^2\lambda ^2)
\end{array}
\right) .
\end{eqnarray}
Here the cloning transformation corresponds to (\ref{tran2}).
Note that the output of copies appear in $a_2, a_3$ qubits.
We have the following four eigenvalues;
\begin{eqnarray}       
&&\frac {1}{3-2\lambda +3\lambda ^2}
\{ \frac {1}{2}(1-6\lambda +\lambda ^2),
~~\frac {1}{2}(1+2\lambda +\lambda ^2),
\nonumber \\
&& 1+\lambda ^2+\frac {1}{2}(1-\lambda )
\sqrt{5+6\lambda +5\lambda ^2},
~~1+\lambda ^2-\frac {1}{2}(1-\lambda )
\sqrt{5+6\lambda +5\lambda ^2}\} .
\label{eigen}
\end{eqnarray}
For optimal phase-covariant quantum cloning,
$\lambda =3-2\sqrt{2}$, the four eigenvalues
are
\begin{eqnarray}
\{ 0, 0, \frac {1}{4}, \frac {3}{4}\} .
\end{eqnarray}
We see that none of the four eigenvalues is negative.
This is different from the UQCM, where one
negative eigenvalue exists for $\lambda =0$.
According to Peres-Horodecki criterion, the copied qubits
in phase-covariant quantum cloning are separable.
Analyzing the four eigenvalues (\ref{eigen}), we find that
the optimal point $\lambda =3-2\sqrt{2}$ is the
only separable point for the copied qubits.
If we analyze the $x-y$ equator, we obtain the same result.

\subsection{Optimal quantum triplicators}
The networks for equatorial qubits can realize the
quantum copying. The copies at the output appear in $a_2$ and $a_3$
qubits. And the output reduced density operator is written as
\begin{eqnarray}
\rho ^{(out)}=\frac {2(1-\lambda ^2)}{3-2\lambda +3\lambda ^2}
\rho ^{(in)}+\frac {1-2\lambda +5\lambda ^2}{6-4\lambda +6\lambda ^2}
\cdot 1.
\label{23out}
\end{eqnarray}
Here, we are also interested in the output state in $a_1$ qubit.
According to the cloning transformations or cloning networks for
equatorial qubits, we find that the reduced density operator
of the output state in $a_1$ qubit can be written as
\begin{eqnarray}
\rho ^{(out)}_{a_1}=\frac {(1+\lambda )^2}{3-2\lambda +3\lambda ^2}
[\rho ^{(in)}]^{T}+\frac {(1-\lambda )^2}{3-2\lambda +3\lambda ^2}
\cdot 1,
\label{1out}
\end{eqnarray}
where the superscript $T$ means transposition.
For $x-z$ equator, the output reduced density operator is invariant
under the action of transposition.
Comparing the output reduced density operators in $a_2$ and $a_3$ qubits
(\ref{23out}) and $a_1$ qubit (\ref{1out}),
in case $\lambda =1/3$, we have a triplicator,
\begin{eqnarray}
\rho ^{(out)}_{a_1}=\rho ^{(out)}_{a_2}=\rho ^{(out)}_{a_3}
=\frac {2}{3}\rho ^{(in)}+\frac {1}{6}\cdot 1,
\end{eqnarray}
with fidelity $\frac {5}{6}$ \cite{BBHB}.
Explicitly, the triplicator
cloning transformation for $x-z$ equator has the form,
\begin{eqnarray}
|0\rangle _{a_1}|00\rangle _{a_2a_3}\rightarrow
\frac {1}{\sqrt{12}}[3|000\rangle _{a_1a_2a_3}+
|011\rangle _{a_1a_2a_3}+|101\rangle _{a_1a_2a_3}+|110\rangle _{a_1a_2a_3}],
\nonumber \\
|1\rangle _{a_1}|00\rangle _{a_2a_3}\rightarrow
\frac {1}{\sqrt{12}}[3|111\rangle _{a_1a_2a_3}+
|100\rangle _{a_1a_2a_3}+|001\rangle _{a_1a_2a_3}+|010\rangle _{a_1a_2a_3}].
\label{xztri}
\end{eqnarray}
For $x-y$ equator, by applying a transformation
$|0\rangle  \leftrightarrow |1\rangle $ in $a_1$ qubit, and still let $\lambda =1/3$,
we find the output density operator in $a_1$ (\ref{1out}) equals to
that of $a_2$ and $a_3$ (\ref{23out}). And the
triplicator cloning for $x-y$ equator takes the form,
\begin{eqnarray}
|0\rangle _{a_1}|00\rangle _{a_2a_3}\rightarrow
\frac {1}{\sqrt{3}}[|001\rangle _{a_1a_2a_3}+|100\rangle _{a_1a_2a_3}+|010\rangle _{a_1a_2a_3}],
\nonumber \\
|1\rangle _{a_1}|00\rangle _{a_2a_3}\rightarrow
\frac {1}{\sqrt{3}}[|110\rangle _{a_1a_2a_3}+|011\rangle _{a_1a_2a_3}+|101\rangle _{a_1a_2a_3}].
\label{xytri}
\end{eqnarray}
The fidelity for quantum triplicator is $\frac {5}{6}$.
Actually, we can find
the fidelity takes the same value $\frac {5}{6}$ when
$\lambda =0$ and $\lambda =1/3$ corresponding to UQCM and
quantum triplicator, respectively.
D'Ariano and Presti \cite{DP} proved that the optimal
fidelity for 1 to 3 phase-covariant quantum cloning is
$\frac {5}{6}$, and presented the cloning
transformation. The quantum triplicators presented
above achieve the bound of the fidelity
and agree with the results in Ref.\cite{DP}.

\section{Optimal 1 to $M$ phase-covariant quantum cloning machines}
We have investigated the $1\rightarrow 2$
and $1\rightarrow 3$ optimal quantum cloning
for equatorial qubits. In what follows,
we shall study the general $N$ to $M$ ($M>N$) phase-covariant
quantum cloning.

We first discuss
$1\rightarrow M$ phase-covariant quantum cloning.
We start from the cloning transformations similar to the
UQCM\cite{GM}, then determine the parameters to
give the highest fidelity, and finally prove that the
determined cloning transformation is the optimal
QCM for equatorial qubits.
For $x-y$ equator
$|\Psi \rangle =(|\uparrow \rangle
+e^{i\phi }|\downarrow \rangle )/\sqrt{2}$,
we suppose the cloning transformations
take the following form,
\begin{eqnarray}
U_{1,M}|\uparrow \rangle \otimes R&=&
\sum _{j=0}^{M-1}\alpha _j|(M-j)\uparrow , j\downarrow \rangle \otimes R_j,
\nonumber \\
U_{1,M}|\downarrow \rangle \otimes R&=&
\sum _{j=0}^{M-1}\alpha _{M-1-j}
|(M-1-j)\uparrow , (j+1)\downarrow \rangle \otimes R_j,
\label{1m}
\end{eqnarray}
where we use the same notations as those of Ref.\cite{GM},
$R$ denotes the initial state of the copy machine and $M-1$
blank copies, $R_j$ are orthogonal normalized states of
ancilla, and $|(M-j)\psi, j)\psi _{\perp}\rangle $ denotes the
symmetric and normalized state with $M-j$ qubits in state
$\psi $ and $j$ qubits in state $\psi _{\perp}$.
For arbitrary input state, the case
$\alpha _j=\sqrt{\frac {2(M-j)}{M(M+1)}}$
is the optimal $1\rightarrow M$ quantum cloning\cite{GM}.
Here we consider the case of $x-y$ equator instead of arbitrary
input state.
The quantum cloning transformations should satisfy
the property of orientation
invariance of the Bloch vector and that we have identical copies. 
The cloning transformation (\ref{1m}) already ensure that
we have $M$ identical copies.
The unitarity of the cloning transformation
demands the relation
$\sum _{j=0}^{M-1}\alpha _j^2=1$.
Under this condition, we can check that the cloning transformation
has the property of orientation invariance of the Bloch vector.
Thus, the relation (\ref{1m}) is the quantum cloning transformation
for $x-y$ equator.
The fidelity of the cloning transformation (\ref{1m}),
takes the form
\begin{eqnarray}
F=\frac {1}{2}[1+\eta (1,M)],
\end{eqnarray}
where
\begin{eqnarray}
\eta (1,M)=\sum _{j=0}^{M-1}\alpha _j\alpha _{M-1-j}
\frac {C_{M-1}^j}{\sqrt{C_M^jC_M^{j+1}}}.
\end{eqnarray}
We examine the cases of $M=2,3$.
For $M=2$, we have
$\alpha _0^2+\alpha _1^2=1$ and 
$\eta (1,M)=\sqrt{2}\alpha _0\alpha _1$. In case
$\alpha _0=\alpha _1=1/\sqrt{2}$, we have the optimal fidelity
and recover the previous result (\ref{xyopt}).
For $M=3$, we have $\alpha _0^2+\alpha _1^2+\alpha _2^2=1$,
and
\begin{eqnarray}
\eta (1,3)=\frac {2}{3}\alpha _1^2
+\frac {2}{\sqrt{3}}\alpha _0\alpha _2.
\end{eqnarray}
For $\alpha _0=\alpha _2=0, \alpha _1=1$, we have
$\eta (1,3)=\frac {2}{3}$, 
which reproduces the case of
quantum triplicator for $x-y$ equator (\ref{xytri}).

We present the result of
1 to $M$ phase-covariant quantum cloning transformations.
When $M$ is even, we have
$\alpha _j=\sqrt{2}/2, j=M/2-1, M/2$ 
and $\alpha _j=0$, otherwise.
When $M$ is odd, we have
$\alpha _j=1, j=(M-1)/2 $ 
and $\alpha _j=0$, otherwise.
The fidelity are
$F=\frac {1}{2}+\frac {\sqrt{M(M+2)}}{4M}$ for M is even,
and $F=\frac {1}{2}+\frac {(M+1)}{4M}$ for M is odd.
The explicit cloning transformations have already been presented in
(\ref{1m}).

Though the fidelity for $M=2, 3$ are optimal, we need to prove that
for general $M$, the fidelity achieve the bound as well.
We apply the same method introduced by Gisin and Massar
in Ref.\cite{GM}.
In order to use some results later, we consider the general
$N$ to $M$ cloning transformation. Generally, we write the $N$ identical
input state for equatorial qubits as
\begin{eqnarray}
|\Psi \rangle ^{\otimes N}=\frac {1}{2^{N/2}}
\sum _{j=0}^Ne^{ij\phi }\sqrt{C_N^j}
|(N-j)\uparrow ,j\downarrow \rangle .
\end{eqnarray}
The most general $N$ to $M$ QCM for equatorial qubits is expressed as
\begin{eqnarray}
|(N-j)\uparrow ,j\downarrow \rangle \otimes R
\rightarrow
\sum _{k=0}^M|(M-k)\uparrow ,k\downarrow \rangle \otimes |R_{jk}\rangle ,
\end{eqnarray}
where $R$ still denotes the $M-N$ blank copies and the initial
state of the QCM, and $|R_{jk}\rangle $ are unnormalized final states of
the ancilla.
The unitarity relation is written as,
\begin{eqnarray}
\sum _{k=0}^M\langle R_{j'k}|R_{jk}\rangle =\delta _{jj'}.
\end{eqnarray}
The fidelity of the QCM takes the form
\begin{eqnarray}      
F=\langle \Psi |\rho ^{out}|\Psi \rangle
=\sum _{j',k',j,k}\langle R_{j'k'}|R_{jk} \rangle A_{j'k'jk},
\end{eqnarray}
where $\rho ^{out}$ is the reduced
density operator of each output qubit by
taking partial trace over all $M$ but one output qubits.
We impose the condition that the output density operator has
the property of Bloch vector invariance, and find the
following for $N=1$,
\begin{eqnarray}
A_{j'k'jk}={1\over 4}\{ \delta _{j'j}\delta _{k'k}
+(1-\delta _{j'j})[\delta _{k',(k+1)}\frac {\sqrt{(M-k)(k+1)}}{M}
+\delta _{k,(k'+1)}\frac {\sqrt{(M-k')(k'+1)}}{M}]\},
\label{amatrix}
\end{eqnarray}
where $j,j'=0, 1$ for case $N=1$.
The optimal fidelity of the QCM for equatorial qubits is
related to the maximal eigenvalue $\lambda _{max}$ of
matrix $A$ by
$F=2\lambda _{max}$ \cite{GM}.
The matrix $A$ (\ref{amatrix}) is a block diagonal
matrix with block $B$ given by
\begin{eqnarray}
B=\frac {1}{4}\left( \begin{array}{cc}
1 & \frac {\sqrt {(M-k)(k+1)}}{M}\\
\frac {\sqrt {(M-k)(k+1)}}{M} & 1\end{array}\right) .
\end{eqnarray}
Thus we have proved that the optimal fidelty
of 1 to $M$ QCM for equatorial qubits takes the form
\begin{eqnarray}
F=2\lambda _{max}=\left\{ \begin{array}{l}
\frac {1}{2}+\frac {\sqrt{M(M+2)}}{4M}, {\rm M~is~even},\\
\frac {1}{2}+\frac {(M+1)}{4M}, {\rm M~is~odd}.\end{array}
\right.
\end{eqnarray}
We thus find in this section the optimal
1 to $M$ phase-covariant quantum cloning transformation,
this is the main result of this paper.

\section{$N$ to $M$ phase-covariant QCM}
We conjecture that the optimal $N$ to $M$ phase-covariant QCM for
$x-y$ equator take the following form:

\noindent 
{\bf Case A}, when $M=N+2L$, the cloning transformation is
\begin{eqnarray}
U_{N,{N+2L}}|(N-j)\uparrow ,j\downarrow \rangle \otimes R
=|(N-j+L)\uparrow ,(j+L)\downarrow \rangle \otimes R_L,
\label{ntoeven}
\end{eqnarray}
which implies that
we just need one ancilla state and can omit it in cloning relation.
The corresponding fidelity is
\begin{eqnarray}
F=\frac {1}{2}+\frac {1}{2^N}
\sum _{j=0}^{N-1}\sqrt{C_N^jC_N^{j+1}}
\frac {\sqrt {(L+j+1)(N+L-j)}}{N+2L}.
\label{ntoeven1}
\end{eqnarray}

\noindent 
{\bf Case B}, when $M=N+2L+1$, the cloning transformation is
\begin{eqnarray}
U_{N,{N+2L+1}}|(N-j)\uparrow ,j\downarrow \rangle \otimes R
&=&\frac {1}{\sqrt{2}}|(N-j+L+1)\uparrow ,(j+L)\downarrow \rangle
\otimes R_L
\noindent \\
&&+\frac {1}{\sqrt{2}}|(N-j+L)\uparrow ,(j+L+1)\downarrow \rangle
\otimes R_{L+1}.
\label{ntoodd}
\end{eqnarray}
The corresponding fidelity is
\begin{eqnarray}
F&=&\frac {1}{2}+\frac {1}{2^{N+1}}
\sum _{j=0}^{N-1}\sqrt{C_N^jC_N^{j+1}}
\nonumber \\
&&\times
\frac {1}{N+2L+1}
[\sqrt {(L+j+1)(N+L-j+1)}+\sqrt {(L+j+2)(N+L-j)}].
\label{ntoodd1}
\end{eqnarray}
When $N=1$, the cloning transformations and the fidelity reduce
to the previous results given in the last section.
For case $N>1$, the upper bound on the fidelity
obtained by the method introduced in Ref.\cite{GM}
is too conservative because that it is sometimes greater
than unity.

It is proved that the optimal fidelity of $N \rightarrow \infty$
quantum cloning equals to the corresponding optimal fidelity of
quantum estimation\cite{BEM,DBE,BCDM}.
In the limit $L \rightarrow \infty $, the fidelity
for $N$ to $N+2L, N+2L+1$ quantum cloning becomes
\begin{eqnarray}
F&=&\frac {1}{2}+\frac {1}{2^{N+1}}
\sum _{j=0}^{N-1}\sqrt{C_N^jC_N^{j+1}}
\end{eqnarray}
which is equal to the optimal fidelity of the quantum
phase-estimation presented in Ref.\cite{DBE}.
This confirms that the optimal fidelity
(\ref{ntoeven1},\ref{ntoodd1}) in the limit
$L \rightarrow \infty$ gives a correct result.
However, we still need a rigorous proof for
the case of general $N$ to $M$ phase-covariant
quantum cloning.

\section{Summary}
In this paper,
the networks consisting of quantum gates for phase-covariant
quantum cloning have been studied.
The copied qubits of phase-covariant cloning machine
are showed to be separable. 
We have given
explicitly the $1\rightarrow M$ cloning transformations for $x-y$
equator. And the optimal fidelity has been proved by using the
method by Gisin and Massar \cite{GM}.
The general $N\rightarrow M$
phase-covariant quantum cloning are conjectured.

\vskip 1truecm
\noindent {\bf Acknowlegements:}
One of the authors HF acknowleges the support of JSPS.
He also would like
to thank Prof.G.Mauro D'Ariano for informing him
their results in Ref.\cite{DP} before it appeared in web.
We thank V.Bu\v{z}ek and D.Bruss for very
useful comments, and we thank N.Gisin and M.Hillery for
communications.

\end{document}